\documentclass[12pt,a4paper]{JHEP3}
\usepackage{amscd,amsmath,amssymb,amsfonts,xspace,mathtools}
\usepackage{xcolor}
\usepackage{tikz-cd}
\usepackage{url}
\let\normalcolor\relax      
\usepackage{latexsym}  
 \usepackage{graphicx}

\usepackage{graphicx, subfig, float}

\usepackage{epsfig,amsfonts,amssymb}
\usepackage{framed}

\newcommand{\refb}[1]{(\ref{#1})}

\newcommand{\nn}{\ensuremath{\nonumber{}}}

\title{
Classical Unattainability of Extremality in non-BPS D-brane Systems 
}

\author{Pranav Kumar, Swapnamay Mondal$^{1}$
\\
{\it $~^1$ Department of Physics, Institute of Science, Banaras Hindu University, Varanasi, 221005, India}
\vspace*{2mm}\\
{\tt pranavkr29@gmail.com, swapno@maths.tcd.ie }
\vspace*{-3mm}
} 

\abstract{
For the worldline theory of an extremal black hole, extremality amounts to vanishing ground state energy. In light of recent gravity results one would expect much like the ground state degeneracy this fine tuned condition too will not be met. It is unclear though whether this should be a quantum artefact or classical. In this paper we consider a non-BPS extremal four-charge Reissner Nordstr\"{o}m black hole in $\mathcal{N}=8$ String theory. It is shown that the microscopic D-brane description fails to admit any extremal state, even classically. This positive energy is expected to destabilize the near horizon $AdS_2$. The positive minimum energy is a direct consequence of the pattern of supersymmetry breaking by the D-branes. The black hole entropy is found to be the logarithm of the number of isolated minima and hence is related to the configurational entropy of the microscopic potential. 
We also find multiple continua of local minima corresponding to marginally bound states of the constituent D-branes.\\

}

\begin{document}
 


\section{Introduction}
The existence of extremal entropy has long been a source of much confusion for the following reason. Extremal black holes carry zero temperature and hence their entropy implies a highly degenerate ground state.  
The appearance of the Planck scale in the denominator in Bekenstein Hawking formula implies the black hole entropy $S_{BH}$ is huge for any macroscopic black hole and $e^{S_{BH}}$ is absolutely humongous. From a quantum systems perspective, the existence of a such a large degeneracy is highly abnormal, unless protected by some symmetry. Supersymmetry would be that symmetry for BPS black holes, but for non-BPS black holes there is a genuine puzzle. This had led many to suspect the very existence of a non-degenerate ground state \cite{Page:2000dk}. In recent years, careful considerations of quantum effects in the gravity side have shown that the low temperature thermodynamics is significantly altered and hint at the existence of the absence of any large ground state degeneracy in the absence of supersymmetry \cite{Chang:2023zqk, Banerjee:2021vjy, Banerjee:2023quv, Mishra:2022gil, Boruch:2022tno, Castro:2021wzn,Nayak:2018qej,Iliesiu:2020qvm}. The difference is made by the boundary modes described by Schwarzian theory \cite{Kit1, Kit2, Maldacena:2016hyu, Maldacena:2016upp, Engelsoy:2016xyb, Stanford:2017thb, Mertens:2017mtv, Lam:2018pvp, Kitaev:2018wpr, Yang:2018gdb, Saad:2019lba, Iliesiu:2019xuh,Choi:2021nnq, Choi:2023syx,Ghosh:2019rcj}. The key features, such as $\log{} T$ dependence of entropy and linear in $T$ energy above extremality, at low temperatures, can qualitatively be explained by assuming the existence of exponentially many low lying states, instead of a degenerate ground state  \cite{Mondal_2025}.

In essence, this suggests black holes are after all like generic quantum many body systems. If so, there is no reason for its ground state energy to be exactly zero. Since the worldvolume Hamiltonian measures the energy above extremality, a non-zero ground state energy amounts to a non-extremal ground state (of an extremal black hole)! So the gravity results indirectly indicate at the non-existence of an extremal ground state for non-BPS black holes.  However it is not clear whether this non-existence is a classical effect or a quantum one. Since classical physics in the near horizon AdS space and the large-N physics of D-brane systems are expected to be holographic duals, and since in gravity side quantum corrections made all the difference, we expect this non-existence should show up in D-brane side even for small charges.

%
%
%

To appreciate the novelty of this phenomenon, let us note that such a situation would seem inconceivable in a field theory set up, as the low energy dynamics of a soliton is pictured as movement in its moduli space. The worldvolume theory thus obtained is bereft of a potential, hence minimum energy in the worldvolume theory vanishes by default. A more sophisticated situation arises for p-brane solutions in supergravity, where microscopic degrees of freedom are not derived from supergravity and does involve non-trivial interactions, but the theory nevertheless admits zero energy vacua as a consequence of supersymmetry \cite{Witten:1995im}. Supersymmetry can be broken by taking intersection of enough number of brane stacks with appropriate orientations \cite{Ortin:1996bz,Mondal:2024qyn}. The resulting extremal non-BPS system  will have a microscopic theory that is not obliged any longer to admit a zero energy vacuum, i.e. an extremal vacuum.

In fact, the most well known microscopic descriptions in String theory are conformal field theories (CFT-s) \cite{Strominger:1996sh, Maldacena:1997de}. Since the central charges of these CFT-s are computed simply by counting bosonic and fermionic degrees of freedom, effects of interactions, if any, might be missed. Given the excellent match with Bekenstein-Hawking entropy it is reasonable to imagine such effects, if any, to be insignificant. Even so, such CFT-s describe the relevant D-brane systems, in a specific regime and hence do not constitute a generic description of these systems. E.g. the CFT accounting for entropy of $D4$ brane black holes in $\mathcal{N}=2$ string theory arises in a certain large volume limit \cite{deBoer:2008fk}. Thus the success of CFT descriptions in accounting for Bekenstein-Hawking entropy does not preclude the importance of interactions.

The question is perhaps best posed for D-brane systems, whose low energy dynamics is that of an effective particle. Such a system would capture the interactions in the form of a potential and the question of classical extremality boils down to existence of zero energy minima. Obviously, the existence of the same is not obvious in absence of supersymmetry. We shall argue in this paper, that for an extremal non-BPS Reissner Nordstr\"{o}m black hole obtained from a BPS black hole by changing signs of some charges, the potential generically does not admit any zero energy minima, implying the non-existence of any extremal state even at classical level! 
We show this explicitly for extremal non-BPS 4-charge Reissner Nordstr\"{o}m black holes carrying pure Ramond Ramond charges in $\mathcal{N}=8$ string theory \cite{Mondal:2024qyn}. 
This phenomenon is of qualitative importance, as positive energy is incompatible with near horizon $AdS_2$ and hence hints that the same should be destabilized. This, together with the non-existence of a degenerate ground state, seems to suggest non-BPS extremal black holes do not really exist.

The paper is organized as follows. In Section \ref{spontext}, we present the general argument for unattainability of extremality for non-BPS extremal D-brane systems. In Section \ref{d2-d2-d2-d6}, discuss the extremal non-BPS four-charge black hole in $\mathcal{N}=8$ string theory in detail. We show that the extremal entropy is related to the configurational entropy. We also discuss various types of minima admitted by the potential. In Section \ref{sdisc}, we discuss various implications of our findings. 

\section{Unattainability of extremality: general argument} \label{spontext}
In this section, we argue that a non-BPS extremal Reissner Nordstr\"{o}m black hole carrying pure Ramond-Ramond charges, such that altering the signs of some of the charges makes the black hole supersymmetric, generically fails to achieve extremality even at a classical level\footnote{We were tempted to use a somewhat more poetic expression ``spontaneous breaking of extremality", in the sense that a theory meant to describe extremal objects fails to admit an extremal ground state. We are thankful to Boris Pioline for pointing out that such a nomenclature would be potentially misleading.}.

Our argument relies on the following well known fact about supersymmetric vacua. Since there are as many F-term equations as there are complex bosons from chiral multiplets, the D-term equations naively seem to impose additional constraints, generically leading to more equations than variables. Thus one may be mistaken to think that supersymmetry is typically spontaneously broken, which is not the case. Let us quickly review how the breaking is avoided.

Consider $N_F$ number of chiral multiplets in some representations of  gauge group of rank $N_G$, e.g. $U(N_G )$. This  results in $N_F$ complex F-term equations and $N_G$ real D-term equations, on $N_F$ complex variables. 
 Due to gauge symmetry, the number of independent real variables is $2N_F - N_G$. On the other hand, the $N_F$ F-term equations are invariant not only under the $U(N_G)$, but also enhanced complexified gauge symmetry $GL(N_G, \mathbb{C})$. Thus the number of independent real F-term equation is  $2(N_F - N_G)$, making the total number of independent real equations $2N_F - N_G$, which is same as the number of independent real variables. 
Note, had there not been the enhanced invariance in the F-term equations, the system would indeed have been over-constrained, generically admitting no solution. The enhanced gauge symmetry in F-term equations is in turn a consequence of holomorphicity. We now argue that for the extremal non-BPS black holes mentioned above, ``F-term" equations of the worldline theory are not holomorphic.


Let us start from the corresponding BPS black hole. Its microscopic description would be given by a supersymmetric quantum mechanics, whose degrees of freedom would be superfields coming from supermultiplets describing the dynamics of the D-branes as well as strings stretched between the D-branes. To get to the non-BPS black hole, one needs to flip the orientations of various D-branes. This does not change the amount of supersymmetry preserved by individual D-brane stacks or even intersection of a few of them, but it will destroy supersymmetry entirely for sufficiently many stacks. 

The key point is that although the amount of supersymmetry preserved by individual or pair of, even triplet of, D-brane stacks do not change, the subgroup of the entire supergroup being preserved does. This effect can be incorporated by appropriate R-symmetry rotations of individual terms. Thus the worldvolume Lagrangian comprises different pieces fashioning same fields arranged in different superfields, hence preserving different supersymmetries, resulting in complete breaking of supersymmetry. 
R-symmetry rotations in particular involve complex conjugation of complex bosons, implying that the F-term equations are no longer holomorphic. Thus the system of F-term and D-term equations is over constrained as per our earlier argument, implying that the potential generically does not admit any zero energy minima.

For the unconvinced reader, in the coming section we provide an explicit example of this phenomenon using the construction of \cite{Mondal:2024qyn}.

\section{Four charge extremal non-BPS black hole in $\mathcal{N}=8$ String theory} \label{d2-d2-d2-d6}

\subsection{D-brane description} 
The black hole in question exists in $\mathcal{N}=8$ supergravity, which is obtained by compactifying type IIA String theory on $T^6$. We shall take the the non-compact spatial coordinates to be  $X_1, X_2, X_3$ and the six compact ones to be $X_4, \dots, X_9$. 
The microscopic description consists of 4 stacks of D-branes, a D2 brane wrapping $X_4 \text{-} X_5$ two cycle,  a D2 brane wrapping $X_6 \text{-} X_7$ two cycle,  a D2 brane wrapping $X_8 \text{-} X_9$ two cycle
and  a D6 brane wrapping the entire six torus. These $X$-s appear as fields on the worldvolume theory of D-branes. From strings stretched between $k^{th}$ and $l^{th}$ D-brane, we get $\mathcal{N}=1$ chiral superfields $Z^{kl}, Z^{kl }$. We shall denote the complex bosons in a chiral superfield by the same symbol as the superfield itself. The meaning would be clear from the context.
For more details on this system, see \cite{Mondal:2024qyn}.

For the purpose of this paper, it suffices to note the potential, which has the following structure
\begin{align}
V &= V_{gauge} + V_D + V_F \, .
\end{align}
The first two pieces are quite straightforward and are given by 
\begin{align}
\nn
V_{gauge} &=  \sum_{k,l=1; k\neq l}^4 \sum_{a=1}^3 \left( X_a^{k}  -  X_a^{l} \right)^2 | Z^{kl} |^2 \, , \\
V_D &= \frac{1}{2} \sum_{k=1}^4 \left( \sum_{l \neq k} | Z^{kl} |^2 -  \sum_{l \neq k} | Z^{lk} |^2 - c^{k}  \right)^2 \,  .
\end{align}
The superscripts are the brane indices. E.g. $X_a^{k}$ is the $X_a$ coordinate of the $k^{th}$ brane.  
The constants $c^{k}$-s are Fayet-Illioloulos parameters and add up to zero. Note $V_{gauge}$ can be made to vanish simply by choosing $X$-s to be coincident, i.e. making the D-branes coincide in $\mathbb{R}^3$. Conversely, if $Z^{kl} =0$ then $X_a^{k} - X_a^{l}$ remain unfixed and the resultant configurations are not quite bound states. As we shall find, many of the minima of the potential are of this sort. 

Coming to $V_F$, we first define 
\begin{align}
\nn
\Phi_1^a &= \frac{1}{\sqrt{2}} \left( X_{4}^a + i \, X_{5}^a   \right) \, , \, \Phi_2^a = \frac{1}{\sqrt{2}} \left( X_{6}^a + i \, X_{7}^a \right) \, , \, \Phi_3^a = \frac{1}{\sqrt{2}} \left( X_{8}^a + i \, X_{9}^a \right) \, , \\
\Phi^{(12)} &:= \Phi^1_3 - \Phi^2_3 \, , \, \Phi^{(23)} := \Phi^2_1 - \Phi^3_1 \, , \, \Phi^{(31)} := \Phi^1_3 - \Phi^2_3 \, .
\end{align}
The F-term potential $V_F$ is then given by
\begin{align}
V_F &= 2 \left( \sum_{i \neq j, i,j =1}^4 |F^{ij}|^2 + \sum_{i \neq j, i,j =1}^4 |G^{ij}|^2 \right) \, ,
\end{align}
where
\begin{align}
\nn
F^{ij} &=  Z^{ij} Z^{ji} + c_{ij} + c'_{ji} \, ,  i,j=1,2,3,4 \, , \\
\nn
G^{21} &=  \Phi^{(12)}  Z^{21} + Z^{23} Z^{31} + Z^{24} . Z^{41}  ~,~ G^{12} =  \Phi^{(12)}  Z^{12} + Z^{13} Z^{32} + Z^{14} . Z^{42}  \, , \\
\nn
G^{31} &=  \Phi^{(31)} Z^{31}  + Z^{32} Z^{21} + ( Z^{43} )^\dagger . ( Z^{14} )^\dagger   ~,~ G^{13} =  \Phi^{(31)}  Z^{13}  + Z^{12} Z^{23} -  (Z^{41})^\dagger . ( Z^{34} )^\dagger \, ,  \\
\nn
G^{32} &=  \Phi^{(23)} Z^{32}  + Z^{31} Z^{12} - i Z^{34} . ( Z^{24})^\dagger  ~,~ G^{23} = \Phi^{(23)} Z^{23} + Z^{21} Z^{13} - i ( Z^{42} )^\dagger . Z^{43}  \, ,  \\
\nn
G^{41} &=   i (\Phi_1^4)^\dagger Z^{41} - i Z^{42} ( Z^{12} )^\dagger - ( Z^{34} )^\dagger ( Z^{13})^\dagger  ~,~ G^{14} =  i Z^{14} (\Phi_1^4)^\dagger - i ( Z^{21} )^\dagger Z^{24} + ( Z^{31} )^\dagger  (Z^{43} )^\dagger \, ,  \\
\nn
G^{42} &=  i (\Phi_2^4)^\dagger Z^{42} - i Z^{43} (Z^{23})^\dagger - i Z^{41} ( Z^{21} )^\dagger  ~,~ G^{24} = i Z^{24} (\Phi_2^4)^\dagger - i (Z^{12} )^\dagger Z^{14} - i ( Z^{32} )^\dagger Z^{34}  \, ,  \\
G^{43} &=  i (\Phi_3^4)^\dagger Z^{43} + (Z^{14})^\dagger ( Z^{31} )^\dagger - ( Z^{24} )^\dagger ( Z^{32} )^\dagger   ~,~ G^{34} =  i Z^{34} ( \Phi_3^4 )^\dagger - ( Z^{13} )^\dagger ( Z^{41} )^\dagger - ( Z^{23} )^\dagger ( Z^{42} )^\dagger  \, . \label{Fterm}
\end{align}
Here $\{ c_{ij}, c'_{ij} \}$ 
are complex non-zero parameters are are related to the vacuum expectation values of background metric and B-fields \cite{Chowdhury:2014yca}. 
In interest of simplicity, we have restricted to the case where each stack contains a single D-brane. Should one wish to put more branes on one or more stacks, various fields will be elevated to vectors/matrices and the worldline theory will become a non-Abelian theory necessitating appropriate modifications which are known \cite{Mondal:2024qyn}.
Now we embark on exploration of the minima of this potential.
\subsection{Non-existence of extremal vacua} 
Classically, an extremal vacuum would correspond to a minima with zero energy in the worldline theory of the black hole. We now argue that such minima do not exist for the potential under consideration.
Although we argue for the brane system with smallest possible charges, i.e a single D-brane per stack, it is straight forward to generalize the argument to higher charges.

For a zero energy minima to exist $V_{gauge}, V_D, V_F$ must all vanish. For $V_{gauge}$ we have noted this can be ensured simply by taking the D-branes to be coinciedent along non-compact directions. As for $V_F$, all the $F$-s and $G$-s in \refb{Fterm} must vanish. The $G$ equations can be used to fix the $\Phi$-s in terms of the $Z$-s. Since each $\Phi$ field is fixed by two equations, these lead to the following consistency conditions:
\begin{align}
\nn
E_{12} &:= Z^{(12)} Z^{(23)} Z^{(31)} + Z^{(12)} Z^{(24)} Z^{(41)}  - Z^{(21)} Z^{(13)} Z^{(32)} - Z^{(21)} Z^{(14)} Z^{(42)} = 0 \, , \\    
\nn
E_{13} &:= Z^{(13)} Z^{(32)} Z^{(21)} + Z^{(13)} (Z^{(43)})^\dagger (Z^{(14)})^\dagger  - Z^{(31)} Z^{(12)} Z^{(23)} + Z^{(31)} (Z^{(41)})^\dagger (Z^{(34)})^\dagger = 0 \, , \\
\nn
E_{23} &:= Z^{(23)} Z^{(31)} Z^{(12)} -i Z^{(23)} Z^{(34)} (Z^{(24)})^\dagger  - Z^{(32)} Z^{(21)} Z^{(13)} + i Z^{(32)} (Z^{(42)})^\dagger Z^{(43)} = 0 \, , \\
\nn
E_{14} &:= i Z^{(14)} Z^{(42)} (Z^{(12)})^\dagger + Z^{(14)} (Z^{(34)})^\dagger (Z^{(13)})^\dagger  - i Z^{(41)} (Z^{(21)})^\dagger Z^{(24)} + Z^{(41)} (Z^{(31)})^\dagger (Z^{(43)})^\dagger = 0 \, , \\
\nn
E_{24} &:= Z^{(24)} Z^{(43)} (Z^{(23)})^\dagger + Z^{(24)} (Z^{(41)})^\dagger (Z^{(21)})^\dagger  - Z^{(42)} (Z^{(12)})^\dagger Z^{(14)} - Z^{(42)} (Z^{(32)})^\dagger Z^{(34)} = 0 \, , \\
E_{34} &:= Z^{(34)} (Z^{(14)})^\dagger (Z^{(31)})^\dagger - Z^{(34)} (Z^{(24)})^\dagger (Z^{(32)})^\dagger + Z^{(43)} (Z^{(13)})^\dagger (Z^{(41)})^\dagger + Z^{(43)} (Z^{(23)})^\dagger (Z^{(42)})^\dagger = 0 \, . \label{Eeqn}
\end{align}
As $F$ equations imply $Z$-s must be non-vanishing, the $E$ equations must admit some non trivial solution in order for a zero energy minima to exist. Alongside gauge invariance, the first equation is also invariant under the scale transformations, altogether having the following $\mathbb{C}^* \times \mathbb{C}^* \times \mathbb{C}^*$ complexified gauge invariance
\begin{align}
Z^{ij} \rightarrow \lambda_i \lambda_j^{-1} Z^{ij} \, , \quad  \lambda_i, \lambda_j \in \mathbb{C}^* \, . \label{comgauge}
\end{align}
But the remaining equations, while gauge invariant, are not invariant under this scale transformation\footnote{Each E equation is invariant under a different scale transformations.} \refb{comgauge}. 
Thus for every pair $(ij)$, we have two equations: $F_{ij} = 0, E_{ij}=0$, totalling 12 complex or 24 real equations. Including 3 D-term equations, and discounting for 3 gauge redundancies, in total we have $24+3-3=24$ real independent equations. 
The number of independent real variables is $24 -3=21$, making the system over-constrained. Hence generically the system would not admit any solutions\footnote{For the supersymmetric case, the six equations analogous to \refb{Eeqn} turn out to be linearly dependent, with only 3 independent. This reduction of 3 complex equations is due to $\mathbb{C}^* \times \mathbb{C}^* \times \mathbb{C}^*$ invariance of F-term equations. This leads to 21 real equations in 21 real variables.}. This is further confirmed numerically in the following.



%
\subsection{Black hole entropy} 
For BPS black holes, the supersymmetric index is protected as one goes from black hole to the D-brane regime. In the absence of supersymmetry, there is no such protected quantity, hence it is not clear what to look for in the D-brane description. The obvious guess, namely zero temperature entropy,  vanishes as the has a unique ground state \cite{Mondal:2024qyn}.
Nevertheless there exist multiple instances of the entropy of a non-BPS black hole being accounted for by a microscopic CFT \cite{Klebanov:1996un, PhysRevLett.77.2368, Breckenridge:1996sn, Maldacena:1996iz, Horowitz:1996ay, Sfetsos:1997xs, Maldacena:1997nx, Maldacena:1996ya, Horowitz:1996ac, Dabholkar:1997rk, Danielsson:2001xe, Cvetic:1996dt, Cvetic:1998vb, Cvetic:1998hg, Das:1997kt, Zhou:1996nz}. Thus we expect some manifestation of black hole entropy to exist in the D-brane system considered. 



Another problem is that we do not know the entropy of the black hole in question from any other computation. This is unlike its supersymmetric cousin, for which the indices were known \cite{Shih:2005qf}. At this point we note that for large charges, a non-BPS extremal black hole will have the same Bekenstein Hawking entropy as its BPS cousin, as they have the same metric. There is no guarantee though that the respective quantum entropy functions would be the same\footnote{We thank Ashoke Sen for discussion on this issue.}. In the absence of any definitive results, it seems the best we can do is to assume that this equivalence holds at least approximately even for small charges. Then we should look for a number in the ballpark of 12, which is the index for the corresponding BPS system \cite{Chowdhury:2014yca}. 


To take a cue from the supersymmetric case, we note that the index computation in a similar setting boils down to counting zero energy potential minima. By supersymmety each minima corresponds to a zero energy ground state, thus this counting amounts to counting degeneracy of zero energy ground states, which is same as the index as all the ground states carry zero angular momentum \cite{Chowdhury:2014yca, Chowdhury:2015gbk}.
In the present case, we have neither zero energy nor degenerate ground states. But due to the complicated nature of the potential, we are still likely to have multiple minima. So we proceed to study these minima in the hope of finding something useful.


For specificity, we work with the following parameters:
\begin{align}
\nn
c_{12} + c'_{12} &= \frac{2}{3}, \, c_{13} + c'_{13} = \frac{3}{5}, \, c_{14} + c'_{14} = \frac{5}{7}, \\
\nn
c_{23} + c'_{23} &= \frac{7}{11}, \, c_{24} + c'_{24} = \frac{11}{13}, \, c_{34} + c'_{34} = \frac{13}{17}, \\
c^1 &= 1, \, c^2 = 2, \, c^3 = 3, \, c^4 = -6 \, . \label{param}
\end{align}
There is a $U(1) \times U(1) \times U(1)$ gauge symmetry, which we fix by taking $Z^{12}, Z^{23}, Z^{34}$ to be real and positive. Our numerical search using tensorflow library \cite{tensorflow2015-whitepaper} for potential minima reveals a rich landscape, containing multiple local minima with curious features. As we go down in the potential landscape, first we encounter multiple continua of local minima, corresponding to marginally bound states of the constituent D-branes. These are discussed  in subection \ref{locmin}. As we go further down, we find multiple isolated minima describing genuine bound states. The isolated minima appear in quadruplets,  due to the following discrete symmetries: 
\begin{enumerate}
\item
A $\mathbb{Z}_2$ symmetry due to sign change of all fields.
\item
A $\mathbb{Z}_2$ symmetry due to sign change of only the following fields: $Z^{(13)}, Z^{(31)}, Z^{(24)}, Z^{(42)}$ and all the $\Phi$ fields.
\end{enumerate}
Note that the second $\mathbb{Z}_2$ is not a gauge symmetry. Similar other $\mathbb{Z}_2$ symmetries, changing signs of some other subsets of $Z$-fields or $\Phi$, are related to this one by gauge transformations. For example consider the symmetry transformation under which $Z^{(12)}, Z^{(21)}, Z^{(34)}, Z^{(43)}$ and all the $\Phi$ fields change sign. This can be achieved by a gauge transformation where $U(1)$-s corresponding to $1^{st}$ and $4^{th}$ brane act by change of signs.
We find 3 such quadruplets with the respective values of the potential: $.6759, 1.1530, 1.1915$. Our criteria for minima is that the norm of the gradient is lower than $10^{-7}$. 
In the following we list out these solutions.
\begin{enumerate}
\item
\underline{$V= .675 \, :$}
\begin{align}
\nn
\Phi^{(12)} &= 0.024 + 	1.517 \, i \, , \, 
\Phi^{(23)} = -0.957   + 	0.436 \, i \, , \,
\Phi^{(31)} = -0.756     + 	 1.508 \, i \, , \\
\nn
\Phi^{(4)}_1 &= 0.393       -  1.387\, i \, , \,
\Phi^{(4)}_2 = -0.103    -  1.230 \, i \, , \,
\Phi^{(4)}_3 = -0.360  - 0.387 \, i \\
\nn
Z^{12} &= 0.851  \, , \,
Z^{13} = -0.086  +     0.302 \, i \, , \,
Z^{14} = -1.438  +    0.931 \, i  \, , \\
\nn
Z^{21} &= -0.826  -  0.031 \, i \, , \,
Z^{23} = 0.503	   \, , \,
Z^{24} = 0.732   +	1.742\, i \, , \\
\nn
Z^{31} &= 0.216  +	1.386 \, i \, , \,
Z^{32} = -1.263  +	0.135 \, i \, , \,
Z^{34} = 0.772 \, , \\
Z^{41} &= 0.273 +     0.248 \, i \, , \,
Z^{42} = -0.164  +    0.446 \, i \, , \,
Z^{43} = -0.873 -  0.001 \, i \, . \label{disc1}
\end{align}
\item
\underline{$V=  1.153 \, :$}
\begin{align}
\nn
\Phi^{(12)} &=  -0.003  -   1.504 \, i \, , \, 
\Phi^{(23)} =  -0.929	-    0.377 \, i \, , \,
\Phi^{(31)} = 0.681	    -    0.107 \, i \, , \\
\nn
\Phi^{(4)}_1 &= 0.317	 +.  1.377	 \, i \, , \,
\Phi^{(4)}_2 =  0.289  +	0.194 \, i \, , \,
\Phi^{(4)}_3 =   -0.482	  +  0.314 \, i \\
\nn
Z^{12} &= 0.755    \, , \,
Z^{13} =  -0.202  -   0.337 \, i \, , \,
Z^{14} =  1.601	+         0.600  \, i  \\
\nn
Z^{21} &=   -0.866   -    0.011   \, i \, , \,
Z^{23} =  0.422 \, , \,
Z^{24} =   -0.367   +	1.864   \, i \\
\nn
Z^{31} &=  0.394   -   1.284    \, i \, , \,
Z^{32} =  -1.291   -   0.305     \, i \, , \,
Z^{34} =  0.728 \, , \\
Z^{41}  &=   -0.354   +	0.157  \, i \, ,\,
Z^{42} =   0.049   +	0.410  \, i \, , \,
Z^{43} =   -0.853   -  0.034 \, i \, . \label{disc2}
\end{align}
\item
\underline{$V=  1.179 \, :$}
\begin{align}
\nn
\Phi^{(12)} &= -0.041 -  1.433 \, i \, , \, 
\Phi^{(23)} = -0.790	-  1.229 \, i \, , \,
\Phi^{(31)} = -0.908	-0.170 \, i \, , \\
\nn
\Phi^{(4)}_1 &= 0.419 + 0.250 \, i \, , \, 
\Phi^{(4)}_2 = -0.235 + 0.384 \, i \, , \, 	
\Phi^{(4)}_3 = -0.547	-0.270 \, i \\
\nn
Z^{12} &= 0.759 \, , \,
 Z^{13} = 0.177	-0.344 \, i \, , \,
Z^{14}  = 1.549 - 0.948 \, i \, , \\ 
\nn
Z^{21}  &= -0.903 + 0.017 \, i \, , \,
Z^{23} = 0.296 \, , \, 	
Z^{24} = 1.087 +  1.645 \, i \, , \\
\nn
Z^{31}  &= -0.631 -  1.297 \, i \, , \,	
Z^{32}  = -1.421 + 0.235 \, i \, , \,
Z^{34} = 0.575 \, , \\
Z^{41} &=-0.342	-  0.217  \, i \, , 
Z^{42}  = -0.181	+   0.331 \, i \, , \,
Z^{43} = -1.094 	 -  0.005 \, i \, . \label{disc3}
\end{align}
\end{enumerate}
We have done several hundreds of runs with our numerical code, with different learning rates, different initial conditions as well as different values of the parameters $c_{ij} + c'_{ij}, c^k$-s. Every time we have got exactly 3 quadruplets with all $Z$-s non-vanishing.  Thus the existence of 12 isolated minima, representing 12 true bound configurations of the D-brane system, seems to be a robust feature. 

Recall that our initial hunch was that we should look for a number in the ballpark of 12, the index for the corresponding BPS system. We are inclined to believe the appearance of exactly the same number of isolated minima as in the BPS case is of some significance. Hence we propose that in the current settings, the logarithm of the number of isolated potential minima should be interpreted as the black hole entropy. Note this notion is consistent with BPS black hole entropy by zero angular momentum conjecture \cite{Chowdhury:2015gbk}.


A related notion of entropy, although not quite the same, appears in energy landscape approach towards complex systems. This approach was first proposed in the context of supercooled liquids \cite{1969Goldstein}. The idea was that at low temperatures the system gets trapped at a valley and long time slow relaxation can be understood as diffusion between different valleys. Later Stillinger and Weber proposed to identify each valley with its minimum, called Inherent Structure \cite{PhysRevA.25.978, Stillinger1995}. In literature Inherent Structures are also known  as basins, pure states, phases, ergodic components etc. Once the isolated minima are thought of as ``states", it is natural to define a ``Configurational Entropy" (also known as Configurational Complexity) $S_c(E) \sim \frac{1}{N} \log{} \Omega(E)$, where $\Omega(E)$ is the number of isolated minima with energy $E$ and $N$ denotes the system size. 
From this perspective the total entropy of the system is thought of as being divided into configurational and ``vibrational entropy". For glassy systems, this approach can be generalized to look at minima of the free energy, corresponding inherent structures being called Free Energy Inherent Structures \cite{coluzzi2003newmethodcomputeconfigurational}. Energy landscape approach is heavily used in molecular science as well \cite{Wales_2004}.

Although $S_c$ captures the spirit of the entropy we need, it differs in details. Firstly we have a point-like system, hence $N$ can be set to $1$. A more serious departure is that $S_c(E)$ clearly depends on energy, whereas our notion of entropy is blind to the energies of the isolated minima. Perhaps this is telling us that after all ``black hole entropy" is not really a natural notion in the D-brane regime, instead it morphs into the unnatural entity 
\begin{align}
S_{brane} &:= \log{} \, ( \# \text{isolated \, minima}) \, . \label{Sbrane}
\end{align}
A more useful way of thinking of the same would be the following. For a closed system while associating entropy with a given energy, one considers states in a narrow energy window about that energy. For any realistic finite system, which will have discreet spectrum without any significant degeneracy, this window can not be taken to be truly infinitesimal \cite{LandauLifshitzStatMech}.
Thus there is an underlying issue of resolution. Coming back to the case at hand, an observer with sufficiently low resolution will count all the minima and would reproduce \refb{Sbrane} as the configurational entropy for lowest possible energy. The morale of the story seems to be that for non-supersymmetric black holes, only some coarse grained information survives in the black holes side.

\subsection{Marginally stable configurations} \label{locmin}
Our numerical exploration suggests the existence of a continuum of local minima for each possible way of disintegration. For each of such continua, the D-brane system breaks up into two or more subsets, such that there are no strings connecting different subsets (i.e. corresponding $Z$ fields vanish) and thus relative positions of these subsets are unfixed, leading to a continuum. Existence of such extrema can be guessed from the observation that the gradient vector of the potential vanishes if all the $Z$ fields are set to zero. Even setting some of the $Z$ fields to zero kills off many components of the gradient vector. It is conceivable the rest of the components might vanish for specific values of remaining fields. This intuition is confirmed by numerical exploration, detailed below. The existence of such minima has previously been noted for the cousin BPS black hole \cite{Kumar:2023kpc}.

We find different classes of potential minima, which could be arranged in the following classes:
\begin{enumerate}
\item
{\bf Unbound configurations:} For the continuum of minima $\mathcal{M}_{(1)(2)(3)(4)}$, all the $Z^{ij}$-s vanish and all the brane separations are unfixed. Like its BPS cousin, it is possible to arrive at a close formula for potential $V_{(1)(2)(3)(4)}$ at  $\mathcal{M}_{(1)(2)(3)(4)}$:
\begin{align}
V_{(1)(2)(3)(4)} &= \frac{1}{2}  \sum_{i=1}^4(c^{i} )^2 + 4 \sum_{(ij)} (c_{ij} + c'_{ij} )^2  \, . \label{vcrit}
\end{align}
For these configurations, all the constituent D-branes roam around freely, unltil they are close enough, in which case unstable directions open up and the system rolls down to lower minima. These configurations have been  proposed to be analogous to giant molecular clouds, which eventually collapse to form stars, which in turn (under certain circumstances) collapse to form black holes in real world \cite{Kumar:2023kpc}.
\item
{\bf Partially bound configurations:} Alongside these, there are minima for which not all but enough $Z^{ij}$-s are vanish, so that that the brane system is a marginal bound state of two or more subsystems. It might be possible to derive analytic expressions for the value of potential for these configurations as well, but we did not attempt it. Our numerical search found continua of local minima for various possible partition of the D-brane system. We are inclined to believe continua corresponding to remaining partitions, namely $(1)(234), \, (2)(134), \, (14)(23)$, also exist, likely to be found upon more intense numerical search.
We list the essential features of all such partially bound configurations in Table \ref{unstmin}, along with the values of the potential for the choice of parameters \refb{param}. We do not see anything illuminating about the precise values of non-vanishing fields at these minima and hence refrain from mentioning the same. As for the value of the potential at the minima, all minima in this class lie below the manifold of unbound configurations $\mathcal{M}_{(1)(2)(3)(4)}$, but above the true bound configurations \refb{disc1}, \refb{disc2}, \refb{disc3}. 
\begin{table}[h]
\begin{center}
\begin{tabular}{|c|c|c|c|c|c|c|c|c|c|c|c||c|}
\hline
manifold & energy & vanishing fields & unfixed fields \\ 
\hline
$\mathcal{M}_{(3)(124)}$ & 11.413 &  $Z^{31}, Z^{13}, Z^{32}, Z^{23}, Z^{34}, Z^{43}$ & $\Phi^{(23)}, \Phi^{(13)}, \Phi^{(4)}_3$  \\
\hline 
$\mathcal{M}_{(4)(123)}$ & 31.243 & $Z^{41}, Z^{14}, Z^{42}, Z^{24}, Z^{43}, Z^{34}$ & $\Phi^{(4)}_1, \Phi^{(4)}_2, \Phi^{(4)}_3$  \\
\hline 
$\mathcal{M}_{(12)(34)}$ & 12.464 & $Z^{13}, Z^{31}, Z^{14}, Z^{41}, Z^{23}, Z^{32}, Z^{24}, Z^{42}$ & $\Phi^{(13)}, \Phi^{(23)}, \Phi^{(4)}_1, \Phi^{(4)}_2$  \\
\hline
$\mathcal{M}_{(13)(24)}$ & 15.777 & $Z^{12}, Z^{21}, Z^{14}, Z^{41}, Z^{23}, Z^{32}, Z^{34}, Z^{43}$ & $\Phi^{(12)}, \Phi^{(23)}, \Phi^{(4)}_1, \Phi^{(4)}_3$  \\
\hline
$\mathcal{M}_{(1)(2)(34)}$ & 14.492 & $Z^{12}, Z^{21}, Z^{13}, Z^{31}, Z^{14}, Z^{41}, Z^{23}, Z^{32}, Z^{24}, Z^{42}$ & $\Phi^{(12)}, \Phi^{(13)}, \Phi^{(23)}, \Phi^{(4)}_1, \Phi^{(4)}_2$  \\
\hline
$\mathcal{M}_{(1)(3)(24)}$ & 18.217 & $Z^{12}, Z^{21}, Z^{13}, Z^{31},  Z^{14}, Z^{41}, Z^{23}, Z^{32}, Z^{34}, Z^{43}$ & $\Phi^{(12)}, \Phi^{(13)}, \Phi^{(23)}, \Phi^{(4)}_1, \Phi^{(4)}_3$  \\
\hline
$\mathcal{M}_{(1)(4)(23)}$ & 35.211 & $Z^{12}, Z^{21}, Z^{13}, Z^{31}, Z^{14}, Z^{41}, Z^{24}, Z^{42}, Z^{43}, Z^{34}$ & $\Phi^{(12)}, \Phi^{(13)}, \Phi^{(4)}_1, \Phi^{(4)}_2, \Phi^{(4)}_3$  \\
\hline 
$\mathcal{M}_{(2)(3)(14)}$ & 22.790 & $Z^{12}, Z^{21}, Z^{13}, Z^{31}, Z^{23}, Z^{32}, Z^{43}, Z^{34}, Z^{24}, Z^{42}$ & $\Phi^{(12)}, \Phi^{(13)}, \Phi^{(23)}, \Phi^{(4)}_2, \Phi^{(4)}_3$  \\
\hline
$\mathcal{M}_{(2)(4)(13)}$ & 34.641 & $Z^{12}, Z^{21}, Z^{14}, Z^{41}, Z^{23}, Z^{32}, Z^{24}, Z^{42}, Z^{34}, Z^{43}$ & $\Phi^{(12)}, \Phi^{(23)}, \Phi^{(4)}_1, \Phi^{(4)}_2, \Phi^{(4)}_3$  \\
\hline
$\mathcal{M}_{(3)(4)(12)}$ & 35.053 & $Z^{13}, Z^{31}, Z^{14}, Z^{41}, Z^{23}, Z^{32}, Z^{24}, Z^{42}, Z^{34}, Z^{43}$ & $\Phi^{(13)}, \Phi^{(23)}, \Phi^{(4)}_1, \Phi^{(4)}_2, \Phi^{(4)}_3$  \\
\hline
\end{tabular}
\end{center}
\caption{List of continua corresponding to partially bound configurations}\label{unstmin}
\end{table}

\end{enumerate}

\section{Discussion} \label{sdisc}
In this paper we have explicitly shown that the microscopic Hamiltonian for four charge extremal non-BPS black holes in $\mathcal{N}=8$ String theory fails to admit any extremal configuration. This argument relies on the broad structure of the potential, hence should hold for other extremal non-BPS black holes realized along similar lines. Furthermore, we retrieve the black hole entropy as the logarithm of the number of isolated minima of the microscopic potential. These altogether seem to suggest that relation between black hole physics and the D-brane physics has to be significantly different for non-BPS black holes as compared to the BPS ones. 

We fall short of realizing the ``exponentially many low lying states" picture \cite{Mondal_2025}, as the minima are not all degenerate. But then we are working with smallest charges. It could be that for large charges different minima indeed come close enough leading to exponentially many low lying states. For the charges at hand, is there a limit where the 12 minima are degenerate? The value of energy at the potential minima is a function of the parameters $c_{ij}, c'_{ij}, c^{k}$, which depend on the background metric and B-field \cite{Chowdhury:2014yca}. We checked numerically that as the values of these parameters are lowered the minima also come down. Since there is no other scale in the system, in the limit where these parameters all vanish, the 12 minima will become degenerate with zero energy. However, so would the marginally stable configurations, creating a continuum of minima. 

Anyhow the appearance of a new notion of entropy that is somewhat similar to configurational entropy of glasses is curious. On one hand this is surprising since black holes and glasses are almost opposite kind of systems: black holes are supposed to be fast scramblers \cite{YasuhiroSekino2008} whereas slow dynamics is characteristic of glasses, black holes are quintessentially thermal systems whereas glassy systems are trapped to an ergodic component for long enough times. 
On the other hand one would be reminded that the phase diagram of Sachdev-Ye model \cite{PhysRevLett.70.3339}, the precursor of SYK model \cite{Kit1, Kit2, Maldacena:2016hyu}, includes a spin glass phase.


A major concern raised by our findings is the fate of the near horizon $AdS_2$. 
If the extremality itself is not genuine, then the existence of the near horizon $AdS_2$ should be in danger. In other words, the positive definite ground state energy is in contradiction with the existence of a near horizon $AdS_2$. Although the $AdS_2$ and positive ground state energy exist in different pictures, i.e. in closed and open string pictures respectively, this does not quite solve the issue.

For supersymmetric black holes, the world line supersymmetric quantum mechanics of the black hole leads to zero energy ground states, separated by a gap from excited states \cite{Chowdhury:2014yca, Chowdhury:2015gbk}. This system flows to a trivial $CFT_1$ with only zero energy states, just like its holographically dual near-horizon $AdS_2$ \cite{Sen:2008vm}. Now that we do not have a similar $CFT_1$ for the non-BPS case, it is unclear what would the putative $AdS_2$ be dual to.
One way out is that $AdS_2$ fails to exist due to its inability to host positive energy excitations. Instabilities of $AdS_2$ space is a much discussed topic which we shall not get into \cite{Strominger1998AdS2, Maldacena:1998uz, Narayan2012DynamicsAdS2}.
 It is not clear whether this suspected instability has anything to do with the Aretakis instability \cite{Aretakis:2011ha, Aretakis:2011hc, Aretakis:2012ei}.

We end this paper with a curious observation: should the branes extend along the flat directions we shall get an $D5 \text{-} D5 \text{-} D5 \text{-} D9$ system, intersecting in $\mathbb{R}^{3}$. The resultant $\mathbb{R}^{3,1}$ will have a positive cosmological constant. This might be useful in constructing a de Sitter vacuum.
\paragraph{Note:} After this work was complete, we came to know about an ongoing work by Abhishek Chowdhury and Sourav Maji, where a different technique is employed to show the absence of zero energy minima of this potential. It is our understanding that their concern is limited to the absence of extremal vacua.

\paragraph{Acknowledgement:}  This work was supported by the Prime Minister Early Career Research Grant ANRF/ECRG/2024/006802/PMS of Anusandhan National Research Foundation. I am indebted to the participants of Indian Strings Meeting 2025, especially Abhishek Chowdhury, Boris Pioline, Ashoke Sen and Sandip Trivedi for useful discussions. 
\newpage

\bibliography{nExt} 
\bibliographystyle{JHEP}   

\end{document}